%% file: CongDraftGM2021Version7.tex
\documentclass[journal]{IEEEtran}

\usepackage[left=0.75in, right=0.75in, top=1in, bottom=0.9in]{geometry}
\usepackage{mathrsfs,amsthm}
\usepackage{cite}

\makeatletter
\def\ps@headings{%
\def\@oddhead{\mbox{}\scriptsize\rightmark \hfil \thepage}%
\def\@evenhead{\scriptsize\thepage \hfil \leftmark\mbox{}}%
\def\@oddfoot{}%
\def\@evenfoot{}}
\makeatother 
\usepackage{graphicx,amsmath,amssymb,stfloats,subfigure,multicol}
\usepackage{epsfig,epsf,psfrag,amssymb,amsfonts,latexsym,graphicx,mathrsfs,subfigure}
\usepackage{bbm}
\usepackage{chngpage}
\usepackage[dvips]{color}
\usepackage{textcomp}
\usepackage{hyperref}
\usepackage{url}
\usepackage[hyphenbreaks]{breakurl}
\usepackage[normalem]{ulem}
\usepackage{algpseudocode}
\newsavebox{\ieeealgbox}

\usepackage{array}

\ifCLASSINFOpdf
\else
\fi
\newtheorem{theorem}{Theorem}

 \def\old#1{}    

\def\RED{{\mbox{\tiny R-ED}}}

\input LTmacros

\begin{document}

\title{Pricing Energy Storage in Real-time Market}
\author{Cong Chen,~\IEEEmembership{Student Member,~IEEE,}
        Lang~Tong,~\IEEEmembership{Fellow,~IEEE}
        and Ye Guo,~\IEEEmembership{Senior Member,~IEEE,}
\thanks{\scriptsize
Cong Chen and Lang Tong (\{cc2662,lt35\}@cornell.edu) are with the School of Electrical and Computer Engineering, Cornell University,  USA. Ye Guo (\url{guo-ye@sz.tsinghua.edu.cn}) is with Tsinghua Berkeley Shenzhen Institute, Shenzhen, P.R. China. This work is supported in part by the National Science Foundation under Award 1809830 and 1932501.}
}

\maketitle



\begin{abstract}
The problem of pricing utility-scale energy storage resources (ESRs) in the real-time electricity market is considered.  Under a rolling-window dispatch model where the operator centrally dispatches generation and consumption under forecasting uncertainty,  it is shown that almost all uniform pricing schemes, including the standard locational marginal pricing (LMP),  result in lost opportunity costs that require out-of-the-market settlements.  It is also shown that such settlements give rise to disincentives for generating firms and storage participants to bid truthfully, even when these market participants are rational price-takers in a competitive market. Temporal locational marginal pricing (TLMP) is proposed for ESRs as a generalization of LMP to an in-market discriminative form.  TLMP is a sum of the system-wide energy price, LMP, and the individual state-of-charge  price.  It is shown that, under arbitrary forecasting errors, the rolling-window implementation of TLMP eliminates the lost opportunity costs and  provides incentives to price-taking firms to bid truthfully with their marginal costs.  Numerical examples show insights into the effects of uniform and non-uniform pricing mechanisms on dispatch following and truthful bidding incentives.
\end{abstract}


\section{Introduction}\label{sec:Intro}

\input{Intro_v5}

\section{Rolling-window dispatch and pricing models}\label{sec:Model}
\input{model_v7}

\section{Dispatch-following incentives and LOC}\label{sec:DispatchFollowLOC}
\input{DispatchFollowLOC_v7}

\section{Truthful-bidding incentives}\label{sec:TruBidIncentive}
\subsection{Bidding strategy of a price-taker}
\input{LOCStorage_v7}

\section{Truthful bidding incentives under TLMP}\label{sec:TLMPStorage}
\input{TLMPStorage_v7}


\section{Case Studies}\label{sec:CaseStudies}
\input{Casestudies_v7}

\section{Conclusions}\label{sec:Conclusion}
This paper considers truthful-bidding  and dispatch-following incentives when pricing utility-scale ESRs under rolling-window dispatch in the real-time wholesale electricity market. We show that almost all uniform pricing schemes need out-of-the-market uplifts and give incentives for generation firms and ESRs to bid untruthfully. It is therefore reasonable to consider nonuniform pricing schemes.   Generalizing the standard LMP to a non-uniform pricing,  TLMP eliminates the need of out-of-the-market settlements and provides incentive compatibility for price-taking firms.

{
\bibliographystyle{IEEEtran}
\bibliography{BIB}
}


\section{Appendix}\label{sec:Appendix}
\input{Proofs_v7}

\input{THMConditionTests_v7}

\end{document}

%% file: LTmacros.tex
\def\beq{\begin{equation}}
\def\eeq{\end{equation}}
\def\bea{\begin{eqnarray}}
\def\eea{\end{eqnarray}}
\def\ba{\begin{array}}
\def\ea{\end{array}}

\def\bitem{\begin{itemize}}
\def\eitem{\end{itemize}}
\def\ben{\begin{enumerate}}
\def\een{\end{enumerate}}

\def\ie{{\it i.e.,\ \/}}

\definecolor{bgrd}{rgb}{1,1,1}
\definecolor{gray}{rgb}{0.5,0.5,0.5}
\definecolor{dkr}{rgb}{0.7,0.1,0.2}
\definecolor{dkb}{rgb}{0.1,0.1,0.8}

\makeatletter
\newdimen{\captionwidth}
\long\def\@makecaption#1#2{%
\captionwidth .9\hsize
\vskip 10pt%
\setbox\@tempboxa\hbox{#1: #2}%
  \ifdim \wd\@tempboxa >\captionwidth%
    \setbox\@tempboxa\hbox{#1:\hspace*{.5em}}%
    \hfil\parbox{\captionwidth}{\raggedright\hangindent \wd\@tempboxa%
    \hangafter=1\unhbox\@tempboxa#2}\hfill%
  \else\centerline{\box\@tempboxa}%
  \fi
}
\makeatother
\def\scalefig#1{\epsfxsize #1\textwidth}

\def\edoc{\end{document}}

\newcommand{\Hmsc}{\mathscr{H}}

\def\zetabf{\hbox{\boldmath$\zeta$\unboldmath}}

\def\lambdabf{\hbox{\boldmath$\lambda$\unboldmath}}

\def\pibf{\hbox{\boldmath$\pi$\unboldmath}}
\def\rhobf{\hbox{\boldmath$\rho$\unboldmath}}

\def\phibf{\hbox{\boldmath$\phi$\unboldmath}}

\def\psibf{\hbox{\boldmath$\psi$\unboldmath}}
\def\omegabf{\hbox{\boldmath$\omega$\unboldmath}}

\def\thetabf{{\mbox{\boldmath$\theta$\unboldmath}}}

\def\ebf{{\bf e}}
\def\fbf{{\bf f}}
\def\gbf{{\bf g}}

\def\pbf{{\bf p}}
\def\qbf{{\bf q}}

\def\Bbf{{\bf B}}

\def\Ebf{{\bf E}}

\def\Gbf{{\bf G}}

\def\Gc{{\cal G}}

\def\Lc{{\cal L}}

%% file: Intro_v5.tex
With the deepening penetration of renewable generation, demand profiles in many electricity markets have shown characteristics of a ``duck-curve", 
 creating frequent and rapid up/down ramp events that pose significant operational challenges.  One way to support ramping is to use multi-interval look-ahead dispatch that anticipates the rise and fall of demand.   Another is to deploy utility-scale storage resources that reduce the need for fast ramps by traditional generators.  In both cases, the dispatch model changes from single-interval to multi-interval dispatch models.

 The standard implementation of a look-ahead dispatch is the {\em rolling-window dispatch,} where the operator uses demand and renewable forecasts in the next few intervals to produce a dispatch plan for the subsequent intervals.  Only the dispatch for the immediate interval is binding. Despite the simplicity and popularity of rolling-window dispatch, pricing generation and consumptions from a rolling-window dispatch is challenging.   Wilson points out in \cite{Wilson:02E} that the rolling-window dispatch and forecasting errors can cause pricing distortions, resulting in undervaluation of generators' intertemporal ramping capabilities.  It is widely recognized that the rolling-window implementation of the standard locational marginal pricing (LMP) imposes  lost opportunity costs (LOC) on generators that have to be compensated by out-of-the-market uplifts.  Such uplifts are discriminative and lack of transparency \cite{Gribik&Hogan&Pope:07, cramton2017ElectricityMarket}.   A significant side effect of such uplifts is that it can create incentives for a generation firm to bid strategically to maximize its profit \cite{cramton2017ElectricityMarket}.

We consider the problem of pricing multi-interval dispatch when utility-scale energy storage resources (ESRs) are part of the real-time electricity market. Under FERC Order 841 \cite{FERC_Storage:841}, ESRs must be able to participate in the market-clearing process as both buyers and sellers, and they are entitled to receive applicable out-of-the-market uplifts. Because it is difficult to audit the actual costs of ESR operations, it is highly desirable that the pricing mechanism is at some level incentive-compatible, which ensures that ESR participants bid truthfully.

Standard LMP does not guarantee incentive compatibility.  Nonetheless,  it is reassuring that price-taking and profit-maximizing firms in a competitive market bid truthfully under the single-interval dispatch and LMP models.   When the dispatch and pricing models change from  single-interval to multiple-interval settings, it is not clear how market participants, especially ESRs, react to the change of incentives.  Will a price-taking ESR participant who bids truthfully under the single-interval LMP model bids strategically to take advantage of the presence of out-of-the-market uplifts?

\subsection{Related Work}
Although incentive issues arising from multi-interval dispatch and pricing models have long been recognized \cite{Wilson:02E,Oren:98Bkchap}, it is only recently that pricing multi-interval dispatch in the real-time market is brought into attention, with emphasis on the lack of dispatch-following incentives in standard rolling-window pricing schemes \cite{CAISO_FRP:15,Schiro:17FERC,Guo&Tong:18Allerton,Hua&etal:19TPS,Zhao&Zheng&Litvinov:19TPS,Guo&Chen&Tong:20arxiv,Chen&Guo&Tong:20TPS}.   A particularly relevant work is \cite{Zhao&Zheng&Litvinov:19TPS} that considers explicitly the participation of  ESRs and the roles of LOC.

The lack of dispatch following incentives can be addressed by providing out-of-the-market uplifts to generators.   The discriminative and non-transparent nature of such uplifts has led to proposals of improvements, including the price-preserving multi-interval pricing (PMP) \cite{hogan2020Multi-intervalPricing}, the constraint-preserving multi-interval pricing (CMP) \cite{Hua&etal:19TPS}, and the Multi-settlement LMP (MLMP) \cite{Schiro:17FERC,Zhao&Zheng&Litvinov:19TPS}.

It turns out that,  at a more fundamental level,  LOC (thus out-of-the-market uplifts) cannot be eliminated by uniform pricing mechanisms \cite{Guo&Chen&Tong:20arxiv}.   The necessity of uplifts under uniform pricing schemes highlights the need to understand ramifications of such uplifts \cite{Gribik&Hogan&Pope:07,Al-Abdullah&Abdi-Khorsand&Hedman:15TPS}, especially from an incentive-compatibility perspective.   To this end, there is little published work in the context of rolling-window dispatch with  ESR participations.   


The temporal locational marginal pricing (TLMP) \cite{Guo&Tong:18Allerton,Guo&Chen&Tong:20arxiv,Chen&Guo&Tong:20TPS} is a nonuniform generalization of LMP that prices generation  based on its contribution to meeting the demand.  Unlike uniform pricing that discriminates generation with out-of-the-market uplifts, TLMP discriminates {\em inside} the market clearing process, therefore eliminating the need of uplifts under arbitrary forecasting error. 

\subsection{Summary of Contributions}
This paper addresses incentive issues when pricing generation and ESRs under the rolling-window dispatch and pricing models.   The main contribution  is twofold.  First, we show that most uniform pricing schemes cannot provide dispatch-following and truthful-bidding incentives for price-taking generators and  ESRs.  The lack of dispatch-following incentives implies that out-of-the-market uplifts are necessary.  We demonstrate through examples that uplift payments result in price-taking ESRs deviating from truthful bidding.

Second, we consider whether a non-uniform (discriminative) pricing mechanism can provide both dispatch-following and truthful-bidding incentives. We show that TLMP, generalized to include ESR participations, not only eliminates the need of out-of-the-market uplifts but also makes truthful-bidding locally optimal for price-taking ESRs.


%% file: model_v7.tex
Consider the rolling-window dispatch of $N$ generators and $M$ ESRs in a single-bus model over a scheduling horizon $\Hmsc=\{1,\cdots, T\}$ of T intervals. With superscript $G$, $D$, and $C$ for generator, ESR discharging power, and charging power, respectively, let $\Gbf_t=\{\Gbf^{\mbox{\tiny G}}_t,\Gbf^{\mbox{\tiny D}}_t,\Gbf^{\mbox{\tiny C}}_t\}$, where $\Gbf^{\mbox{\tiny G}}_t=[\gbf^{\mbox{\tiny G}}[t],\cdots, \gbf^{\mbox{\tiny G}}[t+W-1]] $ be the matrix of all generation variables in the $W$-interval look-ahead window with $\gbf^{\mbox{\tiny G}}[t]=(g^{\mbox{\tiny G}}_{1t},...,g^{\mbox{\tiny G}}_{Nt})$ being the generation vector in interval $t$. Similarly, $\Gbf^{\mbox{\tiny D}}_t$ and $\Gbf^{\mbox{\tiny C}}_t $ represent discharging and charging power of the ESR, respectively.  And $\Ebf_t=[E_{ij}]$ is the matrix of state of charge (SOC) with $E_{ij}$ being the SOC of the $i$th ESR in interval $j$.

The rolling-window dispatch policy, $\Gc^{\RED}$, is defined by a sequence of $W$-interval look-ahead economic dispatch policies $(\Gc_t^{\RED}, t=1, \cdots, T)$. At time $t$,  $\Gc^{\RED}_t$ solves a $W$-interval economic dispatch optimization in interval $\Hmsc_t=\{t,...,t+W-1\}$ using  (i) the realized dispatch $\gbf^{\mbox{\tiny R-ED-G}}[t-1]$ and SOC in interval $t-1$ and (ii) the load forecast $(\hat{d}_t,\cdots,\hat{d}_{t+W-1})$ in $W$ intervals, assuming the forecast of the first interval is accurate, $\hat{d}_t=d_t$. Let $\{f_{nt}^{\mbox{\tiny G}}\}$, $\{f_{it}^{\mbox{\tiny D}}\}$ and $\{f_{it}^{\mbox{\tiny C}}\}$ respectively represent bid-in costs of generator, ESR discharging power, and ESR charging power. The total bid-in cost $F_t(\Gbf_t)$ for the $t$-th rolling-window dispatch can be computed by
 \[
 F_t(\Gbf_t) :=\sum_{t'=t}^{t+W-1} ( \sum_{n=1}^N f^{\mbox{\tiny G}}_{nt'} (g^{\mbox{\tiny G}}_{nt'})+\sum_{i=1}^M (f^{\mbox{\tiny D}}_{it'} (g^{\mbox{\tiny D}}_{it'})- f^{\mbox{\tiny C}}_{it'} (g^{\mbox{\tiny C}}_{it'}))).  
 \]
The rolling-window dispatch policy is defined by
\beq \label{eq:RED}
\begin{array}{lrl}
 & \Gc^{\RED}_t:~\underset{\{\Gbf_t,\Ebf_t\}}{\rm minimize}   &  F_t(\Gbf_t)  \\
&  \mbox{subject to:} & \forall n, \forall i\\ 
& \lambda_{t'}: & \sum_{n=1}^N g^{\mbox{\tiny G}}_{nt'}+\sum_{i=1}^M (g^{\mbox{\tiny D}}_{it'}- g^{\mbox{\tiny C}}_{it'})= \hat{d}_{t'}, \\
&(\underline{\delta }_{it'},\bar{\delta }_{it'}): & \underline{E}_i\le E_{it'} \le \bar{E}_i, \\
& (\underline{\rho}^{\mbox{\tiny G}}_{nt'},\bar{\rho}^{\mbox{\tiny G}}_{nt'}):   & 0 \le g^{\mbox{\tiny G}}_{nt'} \le \bar{g}^{\mbox{\tiny G}}_{n}, \\
&(\underline{\rho}^{\mbox{\tiny D}}_{it'},\bar{\rho}^{\mbox{\tiny D}}_{it'}):   & 0 \le g^{\mbox{\tiny D}}_{it'} \le \bar{g}^{\mbox{\tiny D}}_{i},\\
& (\underline{\rho}^{\mbox{\tiny C}}_{it'},\bar{\rho}^{\mbox{\tiny C}}_{it'}):   & 0 \le g^{\mbox{\tiny C}}_{it'} \le \bar{g}^{\mbox{\tiny C}}_{i}, \\
& (\underline{\mu}_{nt'},\bar{\mu}_{nt'}):  &  -\underline{r}_n\le g^{\mbox{\tiny G}}_{n(t'+1)}-g^{\mbox{\tiny G}}_{nt'} \le \bar{r}_{n},\\
& & \hfill  t \le  t' \le t+W,\\
&\phi_{it'}: &  E_{i(t'-1)}+\xi^{\mbox{\tiny C}}g_{it'}^{\mbox{\tiny C}}-g_{it'}^{\mbox{\tiny D}}/\xi^{\mbox{\tiny D}}=E_{it'}, \\
& & \hfill \mbox{$t+1 \le  t' \le t+W$}.\\
 &  & \mbox{Boundary intertemporal constraints:}\\
 &\phi_{it}: &  E^{\RED}_{i(t-1)}+\xi^{\mbox{\tiny C}}g_{it}^{\mbox{\tiny C}}-g_{it}^{\mbox{\tiny D}}/\xi^{\mbox{\tiny D}}=E_{it}, \\
  & \bar{\mu}_{n(t-1)}:  &  g^{\mbox{\tiny G}}_{nt}-g^{\mbox{\tiny R-ED-G}}_{n(t-1)} \le \bar{r}_{n},\\
  &\underline{\mu}_{n(t-1)}: & g^{\mbox{\tiny R-ED-G}}_{n(t-1)} - g^{\mbox{\tiny G}}_{nt} \le  \underline{r}_n,\\
\end{array}
\eeq
where $\xi^{\mbox{\tiny D}} ,\xi^{\mbox{\tiny C}} \in (0,1]$ are charging and discharging efficiency coefficients. The 3rd to 6th rows are capacity and ramping constraints. 
In (\ref{eq:RED}), the dual variables\footnote{Dual variables for equality constraints are defined in the Lagrangian function ${\cal L}=F_t(\Gbf)-\sum_{t'}\lambda_{t'}(\sum_{n=1}^N g^{\mbox{\tiny G}}_{nt'}+\sum_{i=1}^M (g^{\mbox{\tiny D}}_{it'}- g^{\mbox{\tiny C}}_{it'}) - \hat{d}_{t'})-\sum_{i,t'}\phi_{it'}(E_{i(t'-1)}+\xi^{\mbox{\tiny C}}g_{it'}^{\mbox{\tiny C}}-g_{it'}^{\mbox{\tiny D}}/\xi^{\mbox{\tiny D}}-E_{it'})...$} $(\lambda_{t'},\phi_{it'})$ are respectively associated with the power balance equation and SOC intertemporal equation in interval $t'$, and $(\underline{\delta }_{it'},\bar{\delta }_{it'},\underline{\mu}_{nt'},\bar{\mu}_{nt'},\underline{\rho}^{\mbox{\tiny G}}_{nt'},\bar{\rho}^{\mbox{\tiny G}}_{nt'},\underline{\rho}^{\mbox{\tiny D}}_{it'},\bar{\rho}^{\mbox{\tiny D}}_{it'},\underline{\rho}^{\mbox{\tiny C}}_{it'},\bar{\rho}^{\mbox{\tiny C}}_{it'})$ are dual variables for the lower and upper limits for SOC, ramping and power respectively\footnote{It's assumed throughout the paper that all dual variables for inequality constraints are defined in a way to be always non-negative.}. Let $\xi=\xi^{\mbox{\tiny D}}\xi^{\mbox{\tiny C}}$, by assuming $\frac{\partial}{\partial g^{\mbox{\tiny D}}_{it'}} f^{\mbox{\tiny D}}_{it'} (g^{\mbox{\tiny D}}_{it'}) > \frac{\partial}{\partial g^{\mbox{\tiny C}}_{it'}} f^{{\mbox{\tiny C}}}_{it'} (g^{\mbox{\tiny C}}_{it'})/\xi,\forall i, \forall t'$, constraints $g^{\mbox{\tiny D}}_{it'}g^{\mbox{\tiny C}}_{it'}=0$ can be exactly relaxed in most cases \cite{ZLi2018sufficientStorageRelax}.

Let $(g_{nt}^{{\mbox{\tiny G}}*},g_{it}^{{\mbox{\tiny D}}*},g_{it}^{{\mbox{\tiny C}}*})$ and $(\lambda_{t}^{*})$ be the solution to (\ref{eq:RED}). The rolling-window dispatch, $\Gbf^{\RED}=\{\Gbf^{{\RED}\mbox{\tiny -G}},\Gbf^{{\RED}\mbox{\tiny-D}},\Gbf^{{\RED}\mbox{\tiny-C}}\}$, and rolling-window LMP (R-LMP), $\pibf^{\mbox{\rm\tiny R-LMP}}$, in interval $t$ are given by policy $\Gc^{\RED}$ with
\beq \label{eq:gED}
\begin{array}{c}
g_{nt}^{\RED{\mbox{\tiny-G}}}:= g_{nt}^{{\mbox{\tiny G}}*},~g_{it}^{\RED{\mbox{\tiny-D}}}:= g_{it}^{{\mbox{\tiny D}}*},~g_{it}^{\RED{\mbox{\tiny-C}}}:= g_{it}^{{\mbox{\tiny C}}*},\\
E^{\RED}_{it}=E^*_{it},~\pi^{\mbox{\rm\tiny R-LMP}}_{t} := \lambda_{t}^{*}.
\end{array}
\eeq

%% file: DispatchFollowLOC_v7.tex
\subsection{Lost opportunity cost of ESR}

The LOC payment of individual ESR is a measure of dispatch-following incentives, defined by the difference between the payment that would have been received had the ESR self-scheduled and the payment received within the market clearing process. Under the uniform pricing, let $\pibf$ be the vector of prices over $\Hmsc$. The LOC over the scheduling horizon $\Hmsc$ is given by\footnote{The ESR index is dropped here and thereafter for brevity}
\beq \label{eq:uplift}
\begin{array}{c}
{\rm LOC}(\pibf,\gbf^{\mbox{\tiny R-ED-D}},\gbf^{\mbox{\tiny R-ED-C}})
=Q(\pibf) - \bigg(\pibf^T\gbf^{\mbox{\tiny R-ED-D}}-\pibf^T\gbf^{\mbox{\tiny R-ED-C}}\\
~~~~-\sum_{t=1}^T f^{\mbox{\tiny D}}_{t}(g^{\mbox{\tiny R-ED-D}}_{t})+\sum_{t=1}^T f^{\mbox{\tiny C}}_{t}(g^{\mbox{\tiny R-ED-C}}_{t})\bigg),
\end{array} 
\eeq
where $Q(\pibf)$ is the maximum profit the ESR would have received through the individual profit maximization
\beq \label{eq:Q}
\begin{array}{lcl}
&Q(\pibf)= \underset{\{\pbf^{\mbox{\tiny D}},\pbf^{\mbox{\tiny C}},\ebf\}}{\rm maximize} & \pibf^T\pbf^{\mbox{\tiny D}}-\sum_{t=1}^T f^{\mbox{\tiny D}}_{t}(p^{\mbox{\tiny D}}_{t})\\
&&+\sum_{t=1}^T f^{\mbox{\tiny C}}_{t}(p^{\mbox{\tiny C}}_{t})-\pibf^T\pbf^{\mbox{\tiny C}} \\[0.5em]
& {\rm subject~to}&\\
&\psibf: & \xi^{\mbox{\tiny C}}\pbf^{\mbox{\tiny C}}-\pbf^{\mbox{\tiny D}}/\xi^{\mbox{\tiny D}}=\Bbf\ebf, \\
&(\underline{ \omegabf },\bar{ \omegabf }): & \underline{\Ebf}\le \ebf \le \bar{\Ebf}, \\
 &(\underline{\zetabf}^{\mbox{\tiny D}},\bar{\zetabf}^{\mbox{\tiny D}}):&{\bf 0}\leq \pbf^{\mbox{\tiny D}} \leq\bar{\gbf}^{\mbox{\tiny D}},\\
 &(\underline{\zetabf}^{\mbox{\tiny C}},\bar{\zetabf}^{\mbox{\tiny C}}):&{\bf 0}\leq \pbf^{\mbox{\tiny C}} \leq \bar{\gbf}^{\mbox{\tiny C}},
\end{array} \hfill
\eeq
where $\Bbf$ is $T \times T$ lower bidiagonal matrix with 1 as diagonals and $-1$ left next to diagonals.

\subsection{Conditions for dispatch following incentives}
With the LOC uplift, the revenue of an ESR is maximized for the given price. Therefore, if $\mbox{LOC}=0$, there is no incentive for the ESR to deviate from the dispatch. The following theorem, parallel to Theorem 2 of \cite{Guo&Chen&Tong:20arxiv}, shows that uniform pricing schemes in general cannot eliminate LOC, thus unable to support dispatch-following incentive for ESR. 
\begin{theorem}[Uniform pricing and dispatch-following incentives] \label{thm:StorageUniformPricingLOC}
Under the rolling-window dispatch, there does not exist a uniform pricing scheme under which all ESRs have zero LOC, if there exist ESR $i$ and  $j$ and interval $t^*  \in \Hmsc$  such that
\ben
\item ESR $i$ and  $j$ have distinct bid-in marginal costs for charging and discharging power;
\item both ESRs are ``marginal''  in $t^*$, \ie
\[
 g_{it^*}^{\RED{\mbox{\tiny-D}}} \in (0,\bar{g}^{\mbox{\tiny D}}_i) ~~or ~~g_{it^*}^{\RED{\mbox{\tiny-C}}} \in (0,\bar{g}^{\mbox{\tiny C}}_i),
 \]
 \[
 g_{jt^*}^{\RED{\mbox{\tiny-D}}} \in (0,\bar{g}^{\mbox{\tiny D}}_j) ~~or~~ g_{jt^*}^{\RED{\mbox{\tiny-C}}} \in (0, \bar{g}^{\mbox{\tiny C}}_j);
\]
\item both ESRs don't reach SOC rate limits from $t^*$ to the end of $\Hmsc$.
 \een
\end{theorem}
In our experimental evaluations using practical demand profiles,  we find that these conditions hold with 3.4\% - 85.8\% of time under different parameter settings. So, in corresponding percentage of cases, R-LMP cannot support dispatch-following incentives of ESRs without uplift payments like LOC. See appendix for the proof and simulation.

%% file: LOCStorage_v7.tex


Consider a fixed ESR.
Let $\qbf^{\mbox{\tiny D}}(\cdot) = (q^{\mbox{\tiny D}}_{1}(\cdot), \cdots, q^{\mbox{\tiny D}}_{T}(\cdot))$ be the true cost curve during discharging over $T$ intervals\footnote{The true cost can be evaluated approximately in practice \cite{CAISO_StorageBid:20}.}.  To analyze bidding strategies, we introduce a parametric form of the bid-in cost. Let $\fbf^{\mbox{\tiny D}}(\cdot | \thetabf^{\mbox{\tiny D}}) = (f^{\mbox{\tiny D}}_{t}(\cdot |\theta^{\mbox{\tiny D}}_{t}))$ be the ESR's bid-in cost  (supply) curve parameterized by  $\thetabf^{\mbox{\tiny D}} = (\theta^{\mbox{\tiny D}}_{t})$.  Assume that, at $\thetabf^{\mbox{\tiny D}}=\thetabf^{{\mbox{\tiny D}}*}$, the bid-in cost is the true cost, \ie $\fbf^{\mbox{\tiny D}}(\cdot | \thetabf^{{\mbox{\tiny D}}*})= \qbf^{\mbox{\tiny D}}(\cdot)$. Similarly, $\qbf^{\mbox{\tiny C}}(\cdot),\thetabf^{\mbox{\tiny C}},\fbf^{\mbox{\tiny C}}(\cdot | \thetabf^{\mbox{\tiny C}})$ are defined for ESR's charging operation and $\thetabf=\{\thetabf^{\mbox{\tiny D}},\thetabf^{\mbox{\tiny C}}\}$. For a price-taking ESR, whose bid doesn't affect the market clearing price $\pibf$, the profit is
\beq \label{eq:StoragePi_i}
\begin{array}{ll}
 \Pi(\thetabf) = &\pibf^T\gbf^{\mbox{\tiny R-ED-D}}(\thetabf)-\sum_{t=1}^T q^{\mbox{\tiny D}}_{t}(g^{\mbox{\tiny R-ED-D}}_{t})+\sum_{t=1}^T q^{\mbox{\tiny C}}_{t}(g^{\mbox{\tiny R-ED-C}}_{t})\\
&-\pibf^T\gbf^{\mbox{\tiny R-ED-C}}(\thetabf)+{\rm LOC}(\pibf,\gbf^{\mbox{\tiny R-ED-D}}(\thetabf),\gbf^{\mbox{\tiny R-ED-C}}(\thetabf)),
\end{array}
\eeq
where, $\gbf^{\mbox{\tiny R-ED-D}}(\thetabf),\gbf^{\mbox{\tiny R-ED-C}}(\thetabf)$ are the rolling-window economic dispatch over the entire scheduling horizon when bid-in cost parameter vector is $\thetabf$. The first four of (\ref{eq:StoragePi_i}) are storage surplus, and the last term is LOC computed from (\ref{eq:uplift}). 


A rational ESR maximizes the total profit $ \Pi(\thetabf)$ over bidding strategies parameterized by $\thetabf$. We say that price $\pibf$ is incentive compatible if $\thetabf^*=\{\thetabf^{\mbox{\tiny D}*},\thetabf^{\mbox{\tiny C}*}\}$ corresponding to the true-cost bid is a local maximum of (\ref{eq:StoragePi_i}). 

\subsection{Conditions for truthful-bidding incentives}
We now examine whether there exists a uniform pricing scheme under which a price-taking rational ESR would bid truthfully. Note that, in general, one would expect that a price-taking firm would not bid above its true cost in a competitive market because it will be substituted by another firm who has the same cost bids truthfully.   The intertemporal constraints and the  presence of out-of-the-market uplifts distorts the above argument.  We show below that, for all practical purposes, uniform pricing cannot provide truthful bidding incentives even for price-taking ESRs. 

\begin{theorem} [{Uniform pricing and truthful-bidding incentives}] \label{thm:UntruBidUniformPricingLOC1} 
Assume that the bidding curves of generators and ESRs are linear, and the rolling-window economic dispatch under truthful-biddings, $\Gbf^{\RED}(\thetabf^*)$, is not dual-degenerate. If there exist ESR $i$ and $j$ and interval $t^* \in \Hmsc$ that satisfy the conditions 1)-3) of Theorem \ref{thm:StorageUniformPricingLOC}, then it is suboptimal for ESR i and j to bid truthfully under every uniform price that does not lead to dual degeneracy in (\ref{eq:Q}).
\end{theorem}
 See appendix for the proof. Note that a uniform price that leads to dual degeneracy in (\ref{eq:Q}) must lie in a low dimensional subspace defined by the KKT conditions, which means that uniform prices (almost everywhere) satisfy the non-dual-degeneracy conditions.



%% file: TLMPStorage_v7.tex

\subsection{Temporal locational marginal pricing for ESRs}

Here we extend the temporal locational marginal pricing (TLMP) for ESRs.  As shown in \cite{Guo&Tong:18Allerton,Guo&Chen&Tong:20arxiv}, TLMP is a non-uniform version of LMP defined by the marginal cost of generation and consumption.  For an inelastic demand, the TLMP of consumption is the total cost increase due to the one-MW increase of the demand, which is the same as the standard LMP. For  generators, the price of generation for generator $i$ is defined by the cost reduction due to the one-MW increase of generator $i$'s production.  TLMP for ESRs is similarly defined for charging and discharging separately.

Let $\Gbf^*$  be the solution of the multi-interval economic dispatch in \eqref{eq:RED} and $(\lambda^*_t,\phi^*_t,\underline{\mu}_t^*, \bar{\mu}_t^*,\underline{\delta }_{t}^*,\bar{\delta }_{t}^*,\underline{\rho}_t^*, \bar{\rho}_t^*)$ be the dual variables associated with the constraints in interval $t$. By the envelop theorem, TLMP in interval $t$ for ESR, generators, and demand is defined by
\begin{equation}
\begin{array}{ll}
\mbox{ESR $i$ discharging:} & \pi^{\mbox{\rm\tiny TLMP-D}}_{it} =  \lambda^*_t-1/\xi^{\mbox{\tiny D}} \phi^*_{it},\\
\mbox{ESR $i$ charging:} &  \pi^{\mbox{\rm\tiny TLMP-C}}_{it} =  \lambda^*_t- \xi^{\mbox{\tiny C}}\phi^*_{it},\\
\mbox{Generator $n$:} &   \pi^{\mbox{\rm\tiny TLMP-G}}_{nt} =  \lambda^*_t+ \Delta_{nt}^*,\\
\mbox{Demand:} &   \pi^{\mbox{\rm\tiny TLMP}}_{t} = \lambda_t^*,\\
\end{array}
\end{equation}
where  $\Delta_{nt}^{*} := \Delta \mu_{nt}^{*}-\Delta \mu_{n(t-1)}^{*} $, $\Delta \mu_{nt}^{*}:=\bar{\mu}_{nt}^{*}-\underline{\mu}_{nt}^{*}$, is defined by ramping constraints thus referred to as {\em ramping price}.  Likewise, $\phi_{it}$ is defined by the SOC constraints thus referred to as  {\em SOC price}. The interpretation of TLMP for ESR is therefore the sum of system-wide energy price $\lambda_t$ and SOC price $\phi_{it}^*$ discounted by charging/discharging efficiencies. Likwise, TLMP for generator is the sum of system-wide energy price $\lambda_t$ and the individual ramping price $\Delta_{it}$ \cite{Guo&Chen&Tong:20arxiv}. Similar to the definition of R-LMP in (\ref{eq:gED}), rolling-window TLMP (R-TLMP) is composed of TLMP in the binding interval for each look ahead window.

The intuition behind the TLMP expression can be seen from the Lagrangian of the economic dispatch \eqref{eq:RED} with the optimal multipliers written as
\beq \label{eq:Lagrangian}
\begin{array}{l}
\Lc = \sum_{i,t}  \bigg(f_{it}^{\mbox{\tiny D}}(g^{\mbox{\tiny D}}_{it})- (\lambda_t^* + 1/\xi^{\mbox{\tiny D}}\phi_{it}^*)g_{it}^{\mbox{\tiny D}}  +(\bar{\rho}_{it}^{{\mbox{\tiny D}}*}-\underline{\rho}^{\mbox{\tiny D}*}_{it}) g_{it}^{\mbox{\tiny D}}\bigg)\\
+\sum_{i,t}  \bigg(-f_{it}^{\mbox{\tiny C}}(g^{\mbox{\tiny C}}_{it})
+(\lambda_t^* + \xi^{\mbox{\tiny C}}\phi_{it}^*)g_{it}^{\mbox{\tiny C}} + (\bar{\rho}_{it}^{{\mbox{\tiny C}}*}-\underline{\rho}^{\mbox{\tiny C}*}_{it}) g_{it}^{\mbox{\tiny C}}\bigg) \cdots
\end{array}
\eeq
where the rest of the terms above are independent of $g_{it}^{\mbox{\tiny D}}$ and $g_{it}^{\mbox{\tiny C}}$.  It is evident that, under TLMP, the multi-interval dispatch decouples into single-interval dispatch problems. So intertemporal constraints---SOC constraints ---are decoupled under TLMP, which has significant ramifications in eliminating LOC for ESRs.

The following Theorem establishes that, under TLMP, the LOC for every ESR is zero, and it is locally optimal that every price-taker bids truthfully.  
\begin{theorem}[dispatch and bidding incentives under R-TLMP] \label{thm:TLMPBidLOCESR} For ESR $i$, let $\gbf_i^{\RED \mbox{\tiny -D}},\gbf_i^{\RED \mbox{\tiny-C}}$ be the rolling-window economic dispatch computed from (\ref{eq:gED}) and  $\pibf_i^{\mbox{\tiny D}},\pibf_i^{\mbox{\tiny C}}$ be its R-TLMP.  Then, for all $i$ and under arbitrary demand forecast error,
\beq \label{eq:StorageLOC=0}
\mbox{\rm LOC}(\pibf_i^{\mbox{\tiny D}},\pibf_i^{\mbox{\tiny C}},\gbf_i^{\mbox{\tiny R-ED-D}},\gbf_i^{\mbox{\tiny R-ED-C}})=0,
\eeq
and it is optimal for a price-taking ESR to bid truthfully with its marginal costs of charging and discharging. 
\end{theorem}
See appendix for the proof. Intuitively, from the dispatch following incentive under TLMP, we know $\mbox{\rm LOC}$ in (\ref{eq:StoragePi_i}) is zero and the rolling window dispatch signal, $\ie \gbf_i^{\RED \mbox{\tiny -D}}$ and $\gbf_i^{\RED \mbox{\tiny-C}}$, is an optimal solution for (\ref{eq:Q}). So a truthful-bidding ESR will receive the rolling window dispatch signal optimal for the individual profit maximization, \ie the optimal solution of (\ref{eq:Q}) with $\fbf^{\mbox{\tiny D}}(\cdot | \thetabf^{{\mbox{\tiny D}}*})= \qbf^{\mbox{\tiny D}}(\cdot),\fbf^{\mbox{\tiny C}}(\cdot | \thetabf^{{\mbox{\tiny C}}*})= \qbf^{\mbox{\tiny C}}(\cdot)$ can maximize (\ref{eq:StoragePi_i}) with zero $\mbox{\rm LOC}$.  

%% file: Casestudies_v7.tex
\begin{figure}[h]
\center
\begin{psfrags}
\scalefig{0.25}\epsfbox{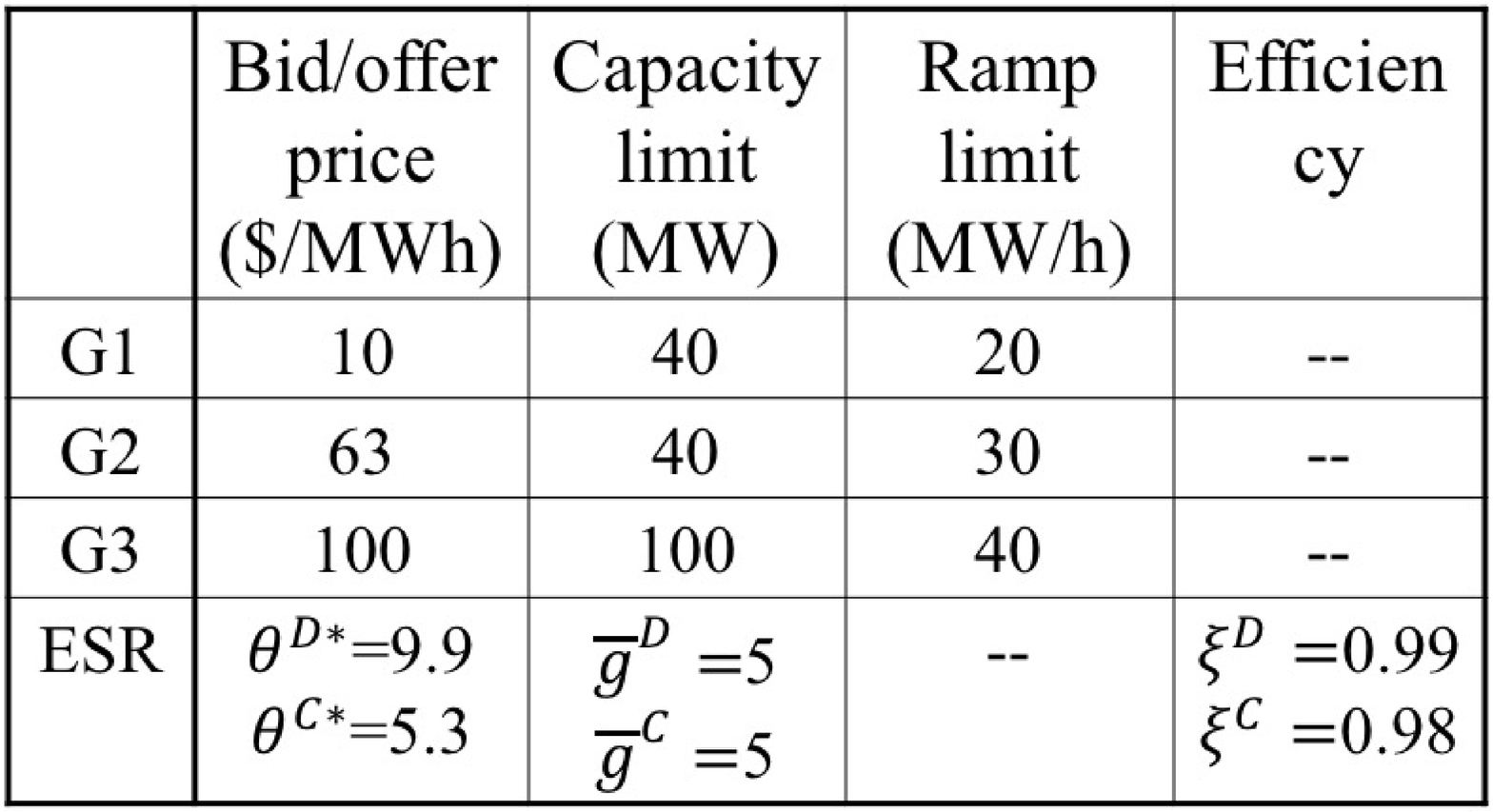}\scalefig{0.22}\epsfbox{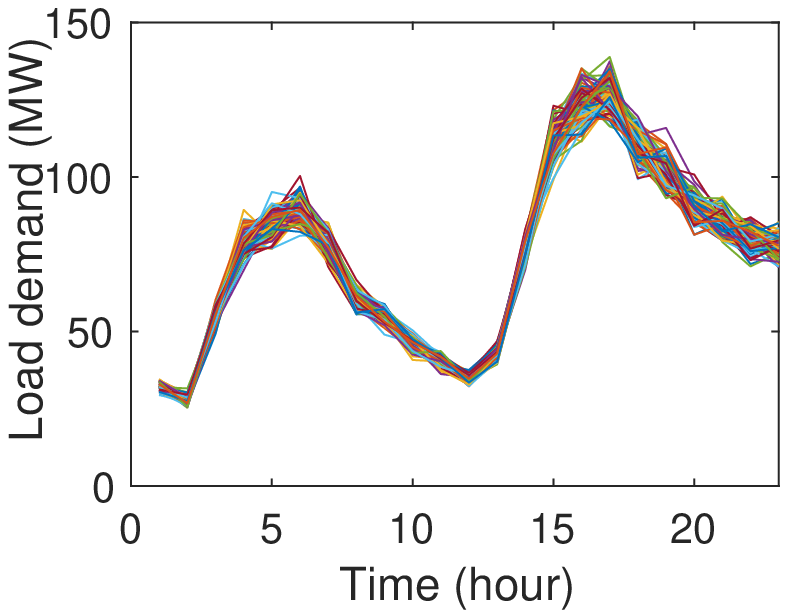}
\end{psfrags}
\vspace{-1em}\caption{\small Left: Parameter Settings. Right: demand traces.}
\label{fig:demand}
\end{figure}

We present here  simulation results involving three generators and one ESR at a single bus to observe performances under multi-interval dispatch and pricing. More simulations about TLMP considering ramping prices of generators are analyzed in \cite{Chen&Guo&Tong:20TPS}. Detailed parameters are shown in Fig~\ref{fig:demand} with minimum generation limits all 0.1 MW and initial SOC 4 MWh. Linear bidding curves were adopted, and we evaluated the performance of benchmark schemes, \ie TLMP and LMP, by varying maximum SOC of ESR.

{\scriptsize
\begin{figure}[h]
\center
\begin{psfrags}
\scalefig{0.17}\epsfbox{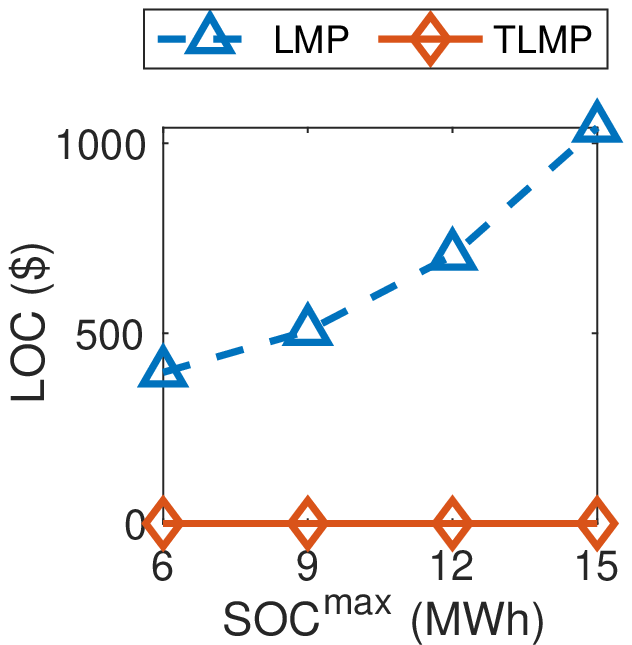}\scalefig{0.17}\epsfbox{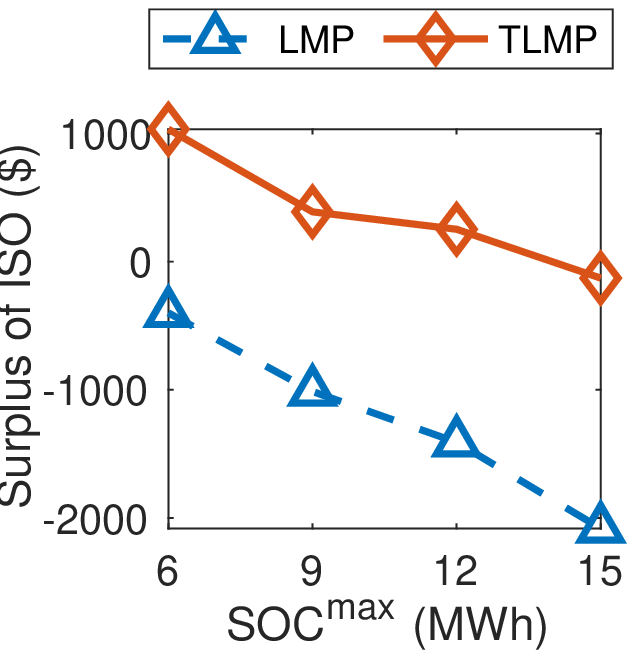}\scalefig{0.17}\epsfbox{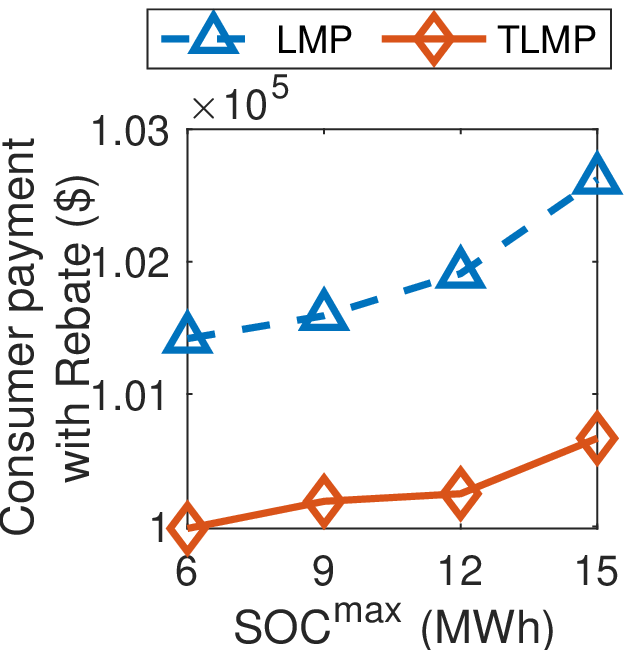}\\
\scalefig{0.17}\epsfbox{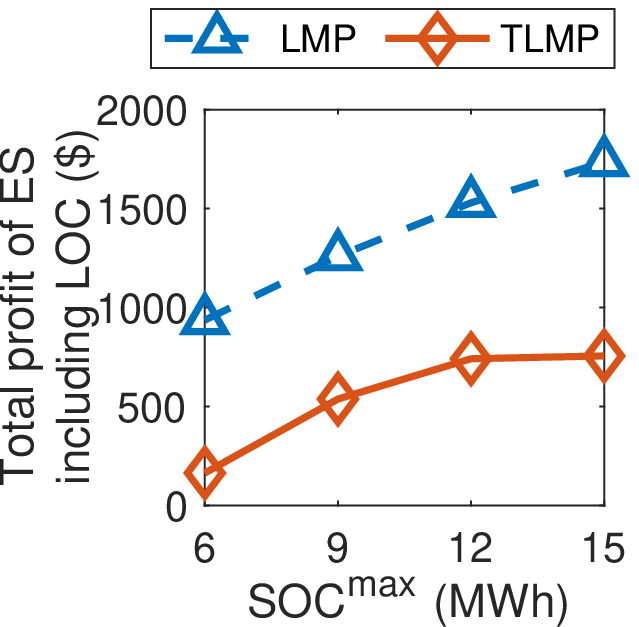}\scalefig{0.17}\epsfbox{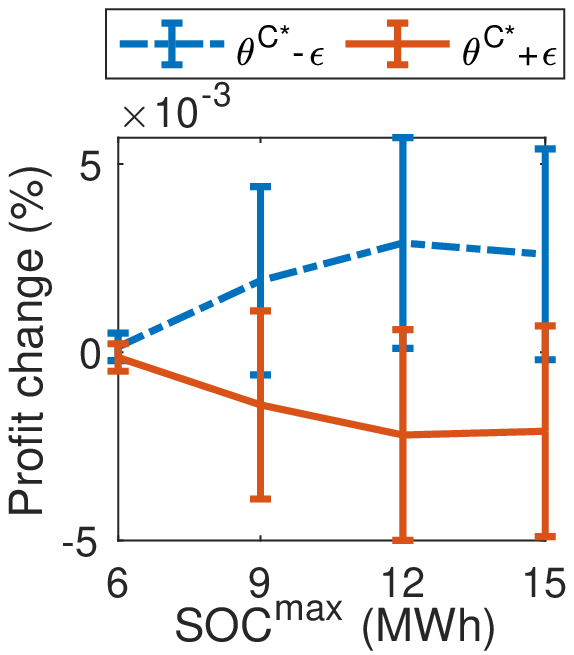}\scalefig{0.17}\epsfbox{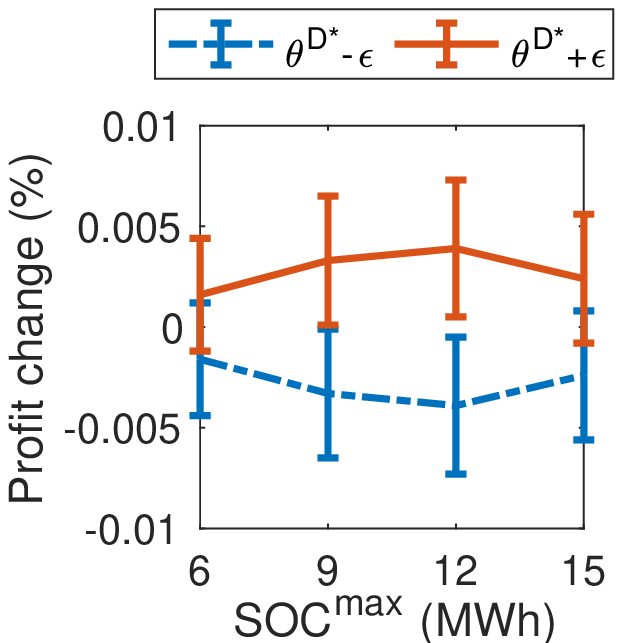}
\end{psfrags}
\vspace{-1.5em}
\caption{\scriptsize Criteria under different maximum SOC. Top Left: LOC. Top Middle: Surplus of ISO. Top Right:  Consumer payment with rebate. Bottom Left: Total profit of ESR. Bottom Middle: Profit change under LMP. Bottom Right:  Profit change under LMP.}
 \label{fig:LOC}
\end{figure}}

The right panel of Fig~\ref{fig:demand} shows the 500 realizations over 24 hour period generated from a CAISO load profile and a standard deviation of $4\%$ of the mean value. We used a standard forecasting error model\footnote{The forecast $\hat{d}_{(t+k)|t}$ at $t$  of demand $d_{t+k}$ is $\hat{d}_{(t+k)|t}=d_{t+k}+\sum_{i=1}^k \epsilon_k$ where $\epsilon_k$ is i.i.d. Gaussian with zero mean and variance $\sigma^2$.} where the demand forecast $\hat{d}_{(t+k)|t}$ of $d_{t+k}$  at time $t$ had error variance $k\sigma^2$ with $\sigma=0.6\%$ for all scenarios. All simulations were conducted with rolling-window optimization over the 24-hour scheduling period, represented by 24 time intervals. And the window size is 4 intervals in each rolling window optimization.

\subsection{Dispatch-following incentives}
Dispatch following incentives are measured by the LOC payment to ESR. A larger LOC payment represents higher incentive to deviate from the dispatch signal (in the absence of a LOC payment). Top left panel of Fig.~\ref{fig:LOC} shows the total LOC payment from the ISO to ESR at different maximum SOC. As predicted by  Theorem \ref{thm:TLMPBidLOCESR}, ESR had strictly zero LOC under TLMP, and positve LOC under LMP. 

\subsection{Surplus of ISO, consumer payments and profit of ESR}
Top middle panel of Fig.~\ref{fig:LOC} shows the ISO's merchandising  surplus including LOC payments. The results validated the fact that uniform pricing schemes, in general, had positive LOC, resulting in a deficit for the ISO. While, coupled with the fact that TLMP always had zero LOC, ISO had a positive merchandising surplus under TLMP in most cases.

We assumed that ISO will redistribute any deficit (and surplus) to the consumers to maintain financially neutral \cite{Schweppe&Caramanis&Tabors&Bohn:88book}. Top right panel of Fig.~\ref{fig:LOC} shows the consumer payments under the assumption, where TLMP was the least expensive for the consumer. Correspondingly, because the operator has zero surplus, ESR had least profit under TLMP, which is shown in the bottom left panel of Fig.~\ref{fig:LOC}. 

\subsection{Truthfully bidding incentives}
In this section, the expected profit changes (including LOC) over different load scenarios were compared when ESR deviated from the truthful marginal cost, $(\theta^{{\mbox{\tiny D}}*},\theta^{{\mbox{\tiny C}}*})=(9.9,5.3)$ \$/MW, by $\varepsilon=0.01$ \$/MW. The bottom middle panel of Fig.~\ref{fig:LOC} shows expectations and standard deviation of the profit changes under the sample average R-LMP. When ESR varied charging bid to $\theta^{{\mbox{\tiny C}}*}-\varepsilon$, ESR had positive profit change on average; while the profit mostly decreased when charging bid was $\theta^{{\mbox{\tiny C}}*}+\varepsilon$. Similar observations can be made on the bottom right panel for discharging bids. So the ESR will decrease charging bids and increase discharging bids to make more profits. However, under R-TLMP, none scenarios was observed to have more profits when ESR deviated from truthful bids.

%% file: Proofs_v7.tex
\subsection{Proof of Theorem~\ref{thm:StorageUniformPricingLOC}}

Suppose the rolling-window economic dispatch, $\{g_{it^*}^{\RED\mbox{\tiny-D}}\}$ and $\{g_{it^*}^{\RED\mbox{\tiny-C}}\}$, is an optimal solution of (\ref{eq:Q}), it must satisfy KKT conditions:
\beq \label{eq:StorageLOCKKT}
\begin{array}{l}
\nabla_{\tiny \pbf^{\mbox{\tiny D}}_{i}} f^{\mbox{\tiny D}}(\pbf^{{\mbox{\tiny D}}*})-\pibf+1/\xi^{\mbox{\tiny D}}\psibf^* + \Delta\zetabf^{{\mbox{\tiny D}}*} ={\bf 0},\\
-\nabla_{\tiny \pbf^{\mbox{\tiny C}}_{i}} f^{\mbox{\tiny C}}(\pbf^{{\mbox{\tiny C}}*})+\pibf-\xi^{\mbox{\tiny C}}\psibf^*+ \Delta\zetabf^{{\mbox{\tiny C}}*} ={\bf 0},
\end{array}
\eeq
where $\Delta\zetabf^{\mbox{\tiny D}*}=\bar{\zetabf}^{\mbox{\tiny D}*} - \underline{\zetabf}^{\mbox{\tiny D}*}$, $\Delta\zetabf^{\mbox{\tiny C}*}=\bar{\zetabf}^{\mbox{\tiny C}*} - \underline{\zetabf}^{\mbox{\tiny C}*}$.

It's known from (\ref{eq:RED}) that $g^{\mbox{\tiny R-ED-D}}_{it'}g^{\mbox{\tiny R-ED-C}}_{it'}=0, \forall i, \forall t'$. So for ESR $i$ and $j$ fulfilling condition 2) and 3) in Theorem~\ref{thm:StorageUniformPricingLOC}, the nonzero charging/discharging power won't reach capacity limits at $t^*$. Meanwhile, SOC won't reach capacity limits from $t^*$ to $T$. And these give $\Delta\zeta^{{\mbox{\tiny D}*}}_{kt^*}=\psi_{kt^*}^*=0, \forall k\in\{i,j\}$. Here we show the contradiction when $g_{it^*}^{\RED{\mbox{\tiny-D}}}  \in (0,\bar{g}^{\mbox{\tiny D}}_i)$ and $g_{jt^*}^{\RED{\mbox{\tiny-D}}} \in (0,\bar{g}^{\mbox{\tiny D}}_j)$. Under the uniform price $\pi_{t^*}$, (\ref{eq:StorageLOCKKT}) gives:
\[
 \pi_{t^*}= \frac{d}{d g} f^{\mbox{\tiny D}}_{it^*}(g_{it^*}^{\RED{\mbox{\tiny-D}}}) = \frac{d}{d g} f^{\mbox{\tiny D}}_{jt^*}(g_{jt^*}^{\RED{\mbox{\tiny-D}}}) 
  \]
This contradicts condition 1). And we can reach similar contradiction when $g_{it^*}^{\RED{\mbox{\tiny-D}}}  \in (0,\bar{g}^{\mbox{\tiny D}}_i)$ and $g_{jt^*}^{\RED{\mbox{\tiny-C}}} \in (0,\bar{g}^{\mbox{\tiny C}}_j)$ , $g_{it^*}^{\RED{\mbox{\tiny-C}}}  \in (0,\bar{g}^{\mbox{\tiny C}}_i)$ and $g_{jt^*}^{\RED{\mbox{\tiny-D}}} \in (0,\bar{g}^{\mbox{\tiny D}}_j)$, and $g_{it^*}^{\RED{\mbox{\tiny-C}}}  \in (0,\bar{g}^{\mbox{\tiny C}}_i)$ and $g_{jt^*}^{\RED{\mbox{\tiny-C}}} \in (0,\bar{g}^{\mbox{\tiny C}}_j)$. So there does not exist a uniform pricing scheme under which both ESR $i$ and $j$ have optimal self-scheduling plans at the rolling-window dispatch signals. And nonzero LOC is needed to compensate ESR.

\subsection{Proof of Theorem~\ref{thm:UntruBidUniformPricingLOC1} }

Known that both ESRs have no dual degeneracy in (\ref{eq:RED}) with linear bidding curve, so for all ESRs, we have $ \frac{\partial}{\partial \theta_{it^*}} g^{\RED \mbox{\tiny -D}}_{kt}(\thetabf^*)=\frac{\partial}{\partial \theta_{it^*}} g^{\RED \mbox{\tiny -C}}_{kt}(\thetabf^*)=0, \forall t \in \Hmsc, \forall i,k \in \{1,...,M\} $. As conditions 1)-3) from Theorem~\ref{thm:StorageUniformPricingLOC} are fulfilled, at least one of ESR $i$ and $j$ has nonzero LOC. Suppose ESR $i$ has nonzero LOC and $g_{it^*}^{\RED{\mbox{\tiny-D}}} \in (0,\bar{g}^{\mbox{\tiny D}}_i)$, with $\pbf^{{\mbox{\tiny D}}*}$ and $\pbf^{{\mbox{\tiny C}}*}$ being the optimal solution for \eqref{eq:Q}, we can rewrite \eqref{eq:StoragePi_i}, the total profit of the ESR $i$, into (\ref{eq:StoragePi_i2}) with nonzero LOC: 

\beq \label{eq:StoragePi_i2}
\begin{array}{ll}
& \Pi_i(\thetabf) =  \pibf^T\gbf_i^{\mbox{\tiny R-ED-D}}(\thetabf)-\sum_{t=1}^T q^{\mbox{\tiny D}}_{it}(g^{\mbox{\tiny R-ED-D}}_{it})\\
 &~~+\sum_{t=1}^T q^{\mbox{\tiny C}}_{it}(g^{\mbox{\tiny R-ED-C}}_{it})-\pibf^T\gbf_i^{\mbox{\tiny R-ED-C}}(\thetabf)\\
 &~~~~+\pibf^T\pbf_i^{{\mbox{\tiny D}}*}-\sum_{t=1}^T f^{\mbox{\tiny D}}_{it}(p^{{\mbox{\tiny D}}*}_{it}|\theta^{\mbox{\tiny D}}_{it})\\
&~~~~~~+\sum_{t=1}^T f^{\mbox{\tiny C}}_{it}(p^{{\mbox{\tiny C}}^*}_{it}|\theta^{\mbox{\tiny C}}_{it})-\pibf^T\pbf_i^{{\mbox{\tiny C}}*}\\
&~~~~~~~~- (\pibf^T\gbf_i^{\mbox{\tiny R-ED-D}}(\thetabf)-\sum_{t=1}^T f^{\mbox{\tiny D}}_{it}(g^{\mbox{\tiny R-ED-D}}_{it}|\theta^{\mbox{\tiny D}}_{it})\\
&~~~~~~~~~~+\sum_{t=1}^T f^{\mbox{\tiny C}}_{it}(g^{\mbox{\tiny R-ED-C}}_{it}|\theta^{\mbox{\tiny C}}_{it})-\pibf^T\gbf_i^{\mbox{\tiny R-ED-C}}(\thetabf)).
\end{array}
\eeq

So, the derivative of \eqref{eq:StoragePi_i2} over $\theta_{it^*}$ can be simplified as
\beq \label{eq:StoragePi_i2Derivative}
\begin{array}{ll}
 \frac{\partial}{\partial \theta^{\mbox{\tiny D}}_{it^*}} \Pi_i(\thetabf^*) = \frac{\partial}{\partial \theta^{\mbox{\tiny D}}_{it^*}} f^{\mbox{\tiny D}}_{it^*}(g^{\mbox{\tiny R-ED-D}}_{it^*}(\thetabf^*)|\theta^{\mbox{\tiny D}}_{it^*})-\frac{\partial}{\partial \theta^{\mbox{\tiny D}}_{it^*}} f^{\mbox{\tiny D}}_{it^*}(p^{{\mbox{\tiny D}}*}_{it^*} |\theta^{\mbox{\tiny D}}_{it^*})\\
~~\sum_{t=1}^T \frac{\partial}{\partial \theta^{\mbox{\tiny D}}_{it^*}} p^{\mbox{\tiny D*}}_{it}(\thetabf)(\pi_t-\frac{\partial}{\partial p^{\mbox{\tiny D}}_{it}} f^{\mbox{\tiny D}}_{it}(p^{\mbox{\tiny D}}_{it^*}(\thetabf^*)|\theta^{\mbox{\tiny D}}_{it^*}))\\
~~~~\sum_{t=1}^T \frac{\partial}{\partial \theta^{\mbox{\tiny D}}_{it^*}} p^{\mbox{\tiny C*}}_{it}(\thetabf)(-\pi_t+\frac{\partial}{\partial p^{\mbox{\tiny C}}_{it}} f^{\mbox{\tiny C}}_{it}(p^{\mbox{\tiny C}}_{it^*}(\thetabf^*)|\theta^{\mbox{\tiny C}}_{it^*})).
\end{array}
\eeq

We know, under every uniform price, ESR $i$ has no dual degeneracy in the individual profit maximization problem (\ref{eq:Q}). So with linear bidding curve, we have $\frac{\partial}{\partial \theta_{it^*}^{\mbox{\tiny-D}}} p^{\mbox{\tiny D*}}_{it}(\thetabf) =  \frac{\partial}{\partial \theta_{it^*}^{\mbox{\tiny D}}} p^{\mbox{\tiny C*}}_{it}(\thetabf) = 0, \forall t \in \Hmsc$. So, only the first two terms of  (\ref{eq:StoragePi_i2Derivative}) are nonzero. Besides, we already know the rolling window dispatch signal is not optimal for (\ref{eq:Q}), $\ie g^{\RED{\mbox{\tiny-D}}}_{it^*}(\thetabf^{*}) \neq p^{{\mbox{\tiny D}}*}_{it^*}$, from Theorem~\ref{thm:StorageUniformPricingLOC}, and we also have $\frac{\partial^2 f^{\mbox{\tiny D}}_{it^*}}{\partial \theta^{\mbox{\tiny D}}_{it^*} \partial g^{\mbox{\tiny D}}_{it^*}}>0$ for the linear bidding curve, so  $\frac{\partial}{\partial \theta^{\mbox{\tiny D}}_{it^*}} \Pi_i(\thetabf^*)   \neq  0$. 

Similarly, we can prove if ESR $i$ has nonzero LOC and $g_{it^*}^{\RED{\mbox{\tiny-C}}} \in (0,\bar{g}^{\mbox{\tiny C}}_i)$ then $\frac{\partial}{\partial \theta^{\mbox{\tiny C}}_{it^*}} \Pi_i(\thetabf^*)   \neq  0$. So, it is suboptimal for ESR $i$ to bid truthfully under every uniform price. 


\subsection{Proof of Theorem~\ref{thm:TLMPBidLOCESR}:} \footnote{Here we focus on ESR $i$ with index $i$ in all variables dropped for brevity.}
KKT condition to (\ref{eq:RED}) is
\beq \label{eq:DispatchKKT}
\begin{array}{l}
\nabla_{\tiny \gbf^{\mbox{\tiny D}}_{i}} f^{\mbox{\tiny D}}(\gbf^{{\mbox{\tiny D}}*})-\lambdabf+1/\xi^{\mbox{\tiny D}}\phibf^* + \Delta\rhobf^{{\mbox{\tiny D}}*} ={\bf 0},\\
-\nabla_{\tiny \gbf^{\mbox{\tiny C}}_{i}} f^{\mbox{\tiny C}}(\gbf^{{\mbox{\tiny C}}*})+\lambdabf-\xi^{\mbox{\tiny C}}\phibf^* + \Delta\rhobf^{{\mbox{\tiny C}}*} ={\bf 0},
\end{array}
\eeq
where $\Delta\rhobf^{\mbox{\tiny D}*}=\bar{\rhobf}^{\mbox{\tiny D}*} - \underline{\rhobf}^{\mbox{\tiny D}*}$, $\Delta\rhobf^{\mbox{\tiny C}*}=\bar{\rhobf}^{\mbox{\tiny C}*} - \underline{\rhobf}^{\mbox{\tiny C}*}$. Given rolling-window dispatch signals $\gbf^{\RED \mbox{\tiny -D}},\gbf^{\RED \mbox{\tiny-C}}$ and R-TLMP, $\pibf^{\mbox{\tiny D}}$ and $\pibf^{\mbox{\tiny C}}$, the KKT conditions of the individual optimization shown in (\ref{eq:StorageLOCKKT}) can be satisfied like equation (\ref{eq:DispatchKKT}) by setting $\psibf^*=\mathbf{0},\gbf^{\RED{\mbox{\tiny-D}}}= \pbf^{{\mbox{\tiny D}}*},\gbf^{\RED{\mbox{\tiny-C}}}= \pbf^{{\mbox{\tiny C}}*},\Delta\rhobf^{\mbox{\tiny D}*}=\Delta\zetabf^{\mbox{\tiny D}*},\Delta\rhobf^{\mbox{\tiny C}*}=\Delta\zetabf^{\mbox{\tiny C}*}$. So TLMP can support dispatch-following incentive of ESRs with zero LOC.

Next we show the truthful bidding incentives. We know that TLMP can always give $\mbox{\rm LOC}=0$. So under the price-taker assumption, from (\ref{eq:StoragePi_i}), we have
\beq \label{eq:StoragePi_TLMP}
\begin{array}{lcl}
&&\Pi(\thetabf^*) = (\pibf^{\mbox{\tiny D}})^T\gbf^{\mbox{\tiny R-ED-D}}(\thetabf^*)-\sum_{t=1}^T q^{\mbox{\tiny D}}_{t}(g^{\mbox{\tiny R-ED-D}}_{t}(\thetabf^*))\\
&&~~+\sum_{t=1}^T q^{\mbox{\tiny C}}_{t}(g^{\mbox{\tiny R-ED-C}}_{t}(\thetabf^*))-(\pibf^{\mbox{\tiny C}})^T\gbf^{\mbox{\tiny R-ED-C}}(\thetabf^*)\\
&&~~~~ \ge (\pibf^{\mbox{\tiny D}})^T\gbf^{\mbox{\tiny D}}-\sum_{t=1}^T q^{\mbox{\tiny D}}_{t}(g^{\mbox{\tiny D}}_{t})\\
&&~~~~~~+\sum_{t=1}^T q^{\mbox{\tiny C}}_{t}(g^{\mbox{\tiny C}}_{t})-(\pibf^{\mbox{\tiny C}})^T\gbf^{\mbox{\tiny C}},
\end{array}
\eeq
 for every $\gbf^{\mbox{\tiny D}},\gbf^{\mbox{\tiny C}}$ in the profit maximization problem, which is (\ref{eq:Q}) with $\fbf^{\mbox{\tiny D}}(\cdot | \thetabf^{{\mbox{\tiny D}}*})= \qbf^{\mbox{\tiny D}}(\cdot),\fbf^{\mbox{\tiny C}}(\cdot | \thetabf^{{\mbox{\tiny C}}*})= \qbf^{\mbox{\tiny C}}(\cdot)$.  Because a price-taker's bid under TLMP can only influence dispatch $\gbf^{\mbox{\tiny R-ED-D}}(\thetabf)$ and $\gbf^{\mbox{\tiny R-ED-C}}(\thetabf)$, we have $\Pi(\thetabf^*) \ge \Pi(\thetabf)$. \hfil

%% file: THMConditionTests_v7.tex
\subsection{Simulations on the conditions in Theorem~\ref{thm:StorageUniformPricingLOC}}
\begin{figure}[h]
\center
\begin{psfrags}
\scalefig{0.3}\epsfbox{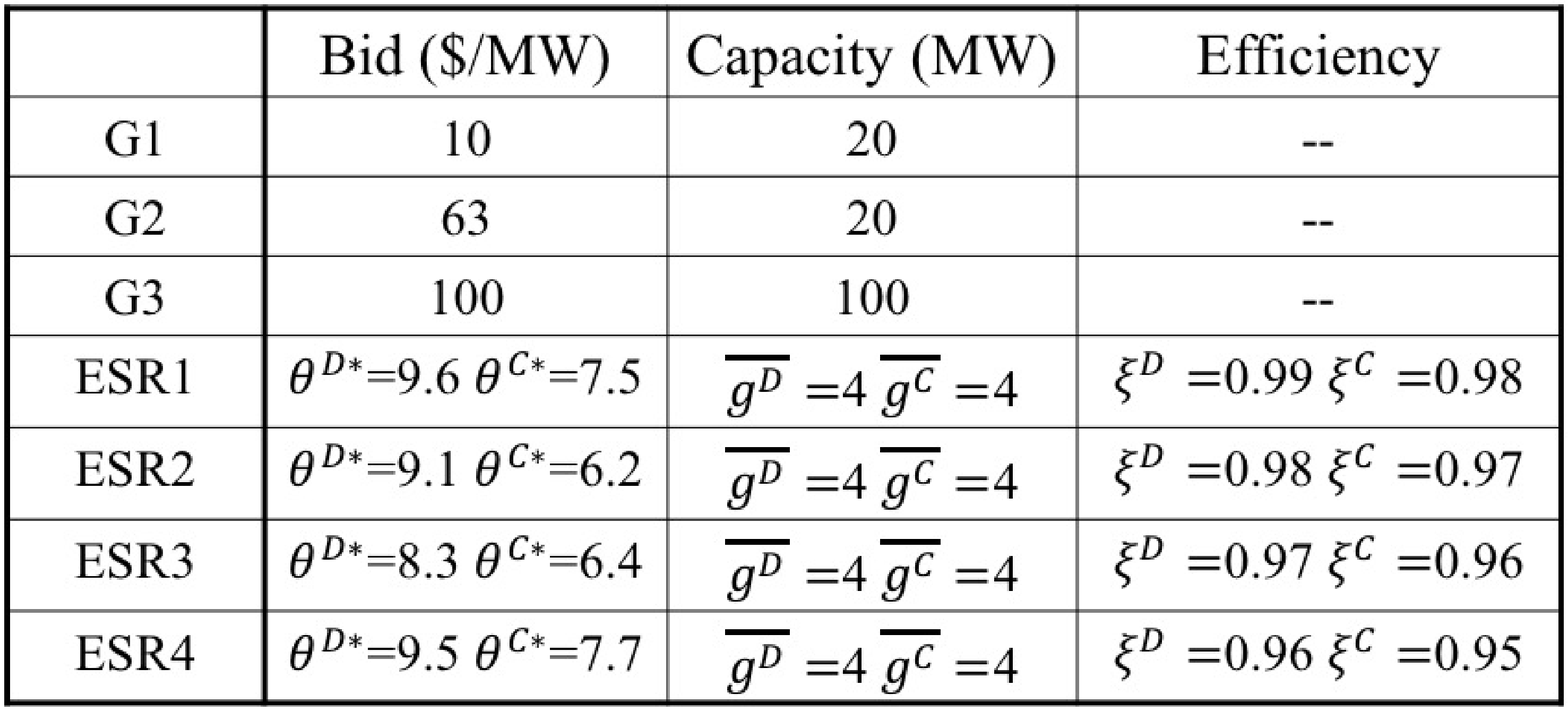}\scalefig{0.20}\epsfbox{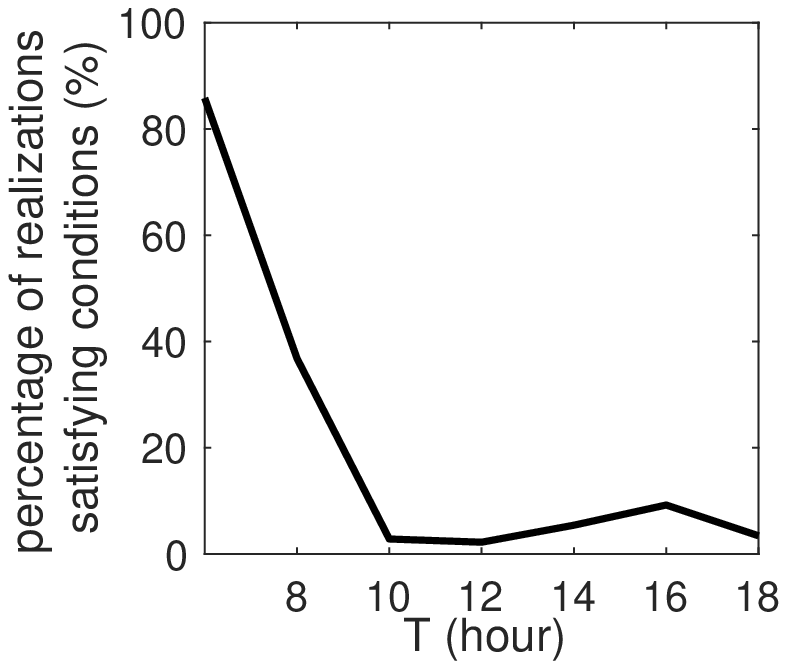}
\end{psfrags}
\vspace{-1em}\caption{\small Left: Parameter Settings. Right: Percentage of realizations satisfying conditions vs. $T$.}
\label{fig:SettingResult}
\end{figure}
We present empirical test results on how  frequently conditions  in Theorem~\ref{thm:StorageUniformPricingLOC} hold. Parameters of this single node case are shown in Fig~\ref{fig:SettingResult} with minimum generation limits all 0 MW and initial SOC 4 MWh. Linear bidding curves were adopted, and we evaluated the percentage of realizations satisfying conditions by varying the sizes $T$ of the rolling-window scheduling horizons, $\Hmsc=\{1,\cdots, T\}$. The average load scenario and the way to generate random load realizations were the same as Section \ref{sec:CaseStudies} and 500 realizations were tested with rolling window size four-interval. It can be observed in the right panel of Fig.~\ref{fig:SettingResult} that 3.4\% - 85.8\% realizations satisfied the conditions given in Theorem~\ref{thm:StorageUniformPricingLOC} under different horizons of the rolling-window dispatch.